\documentclass[runningheads]{llncs}
\usepackage{graphicx}
\usepackage{dcolumn}
\usepackage{multirow}
\graphicspath{{all_charts_and_graphics/}}
\usepackage{wrapfig}

\usepackage{caption}
\usepackage{amsmath}
\usepackage[hidelinks]{hyperref}
\usepackage[table,xcdraw]{xcolor}

\usepackage{algorithm}
\usepackage{algpseudocode}
\usepackage{graphicx}
\begin{document}
{\Large\bfseries \title{Movie Recommendation System using Composite Ranking}}
%
%
\author{Irish Mehta\orcidID{0000-0002-7001-258X} \and
Aashal Kamdar\orcidID{0000-0002-7067-425X}}
%
\authorrunning{Irish Mehta and Aashal Kamdar}
%
\institute{
Birla Institute of Technology \& Science Pilani, K.K. Birla Goa Campus, India 
\email{irish.mehta@gmail.com} \and
Manipal Institute of Technology, Karnataka, India \email{aashalkamdar@gmail.com} 
}
\maketitle              
\begin{abstract}
In today's world, abundant digital content like e-books, movies, videos and articles are available for consumption. It is daunting to review everything accessible and decide what to watch next. Consequently, digital media providers want to capitalise on this confusion and tackle it to increase user engagement, eventually leading to higher revenues. Content providers often utilise recommendation systems as an efficacious approach for combating such information overload. This paper concentrates on developing a synthetic approach for recommending movies. Traditionally, movie recommendation systems use either collaborative filtering, which utilises user interaction with the media, or content-based filtering, which makes use of the movie’s available metadata. Technological advancements have also introduced a hybrid technique that integrates both systems. However, our approach deals solely with content-based recommendations, further enhancing it with a ranking algorithm based on content similarity metrics. The three metrics contributing to the ranking are similarity in metadata, visual content, and user reviews of the movies. We use text vectorization followed by cosine similarity for metadata, feature extraction by a pre-trained VGG19 followed by K-means clustering for visual content, and a comparison of sentiments for user reviews. Such a system allows viewers to know movies that "feel" the same. 

\keywords{Recommendation systems  \and Content-based filtering \and Sentiment Analysis \and Visual Similarity.}
\end{abstract}
\section{Introduction}
The internet has become widespread, leading to an unlikely problem for users; the problem of having too many choices. From choosing electronics like mobiles and laptops to choosing a university for graduation, there is just an exorbitant amount of information at the tip of our fingers. This might get overwhelming, so recommendation systems were introduced as an effective method of combating this information overload. Even though the idea of recommendation systems is relatively new in the field of research, it has long been an integral component of society, significantly impacting our lives and also those around us\\  \cite{1-intro}.  

Researchers have developed various recommender systems up to this point for many different types of industries. Recommender systems benefit both service providers and users \cite{2-intro}. They help companies with customer retention, thus increasing revenues and reducing the time spent by a user looking for the next best item. There are mainly three different types of recommendation systems that are utilised, which have been described below:

\textbf{Content-Based Filtering:}
Such systems are based on showing the user more content that is similar to what they already liked or have liked in the past\\  \cite{1-intro}. The similarity between two objects can be estimated based on their related features. This method uses features and likes provided by the user in order to curate recommendations that the user might like.

\textbf{Collaborative Filtering:}
This method focuses on finding similar users and recommends what similar users like. This technique can filter out items that a user might enjoy basis ratings or comments of other users \cite{9-lit}. 

\textbf{Hybrid Recommender Systems:}
The idea behind this method is to provide a combination of two recommendation systems so as to overcome the shortcomings of each system \cite{1-intro}.

In this paper, we focus on content-based recommendation systems and aim to improve the recommendations based on a composite ranking system involving a combination of visual aspects and user reviews of the content. We first use a metadata-based recommendation system to get a set of initial recommendations for movies. To understand the visual features of the reference and recommended movies, we utilise key frames of the movie trailers, VGG19 for extraction of features, and K Means clustering for grouping similar key frames. This is followed by a novel approach to compare the closeness of both the trailers. Additionally, we use sentiment analysis to understand how the wider audience has received it. Ultimately, both of these are combined to create a ranking algorithm.

\section{Related Work}
At Duke University, recommender systems developed into a separate field of study in the middle of the 1970s. The first recommendation system was developed at the Xerox Palo Alto Research Centre and was called Tapestry \cite{2-lit}. It was created as a solution to the rising use of electronic mail, which led to a massive influx of documents. 

Recommendation systems have been defined as a decision-making method for users in complicated information settings \cite{24-lit}. They can also be defined as a tool for assisting and enhancing the social process of using recommendations from others to make decisions when one lacks sufficient personal knowledge or expertise of other options \cite{25-lit}.  User information overload is addressed by recommender systems by offering users individualised, exclusive content and service recommendations \cite{2-intro}.

Several methods have been developed for building recommendation systems that employ collaborative filtering, content-based filtering and hybrid filtering. The most popular personalised recommendation technique in use is the collaborative filtering algorithm \cite{9-lit}. Collaborative filtering gained widespread attention when Amazon.com’s research team published how the company uses collaborative filtering to improve its user recommendations \cite{10-lit}. According to them, Amazon utilised a memory-based approach by using an item-to-item matrix for similar items. Content-based filtering provides recommendations to a user by correlating the information describing an item with other items in the database. Simon Philip et al. \cite{11-lit} deployed a content-based recommendation system for recommending research papers for a digital library. Robin van Meteran et al. \cite{26-lit} use content-based filtering to suggest small articles on home improvements. 

Nevertheless, these methods also have their limitations. Limited content analysis, overspecialization, and data sparsity are some issues with content-based filtering strategies \cite{12-lit}.
Additionally, cold-start, sparsity, and scalability issues are present in collaborative techniques. These can be mitigated by creating a system that combines the features of different filtering techniques to provide increased accuracy. This method is known as hybrid filtering \cite{13-lit}. Mohammed Baidad et al. \cite{15-lit} used a hybrid recommendation system to improve the quality of recommendations in education to suit the needs of every learner since everyone learns differently the majority of the time. By merging various matrix manipulation techniques with fundamental recommendation strategies, hybrid filtering attempts to tackle the data sparsity and cold-start problems. They also strive to make better use of product attributes, product testimonials, user demographic information, or other well-known user traits \cite{27-lit}. Y Dang et al. \cite{28-lit} propose a hybrid collaborative filtering algorithm for the recommendation of news and interesting information sources. Konstas et al. \cite{29-lit} propose a music recommendation system that incorporates play counts, tagging data, and social relationships.

Yashar et al. have developed a recommendation system based on the visual features of the content itself \cite{20-lit}. Their method focuses on how a content-based recommendation system can give a better recommendation based on the visual similarity of the two movies based on the theory of Applied Media Aesthetics. The team at MediaFutures has developed multiple recommendation systems incorporating Deep Content Features, i.e., visual features, to solve the cold start problem and provide a comparatively better recommendation than metadata \cite{21-lit}.

Elham Asani et al. \cite{22-lit} developed a recommendation system to suggest restaurants to users by eliciting their food preferences from their comments. Anmol Chauhan et al. \cite{23-lit} use sentiment analysis to recommend movies to a user based on their view history.

However, almost all the existing recommendation systems utilise visual similarity for either standalone recommendations or to recommend in the case of a cold start problem. Regarding our contribution, there is no other algorithm designed to rank recommendation systems in three dimensions of the content, i.e., metadata, visual and sentiment analysis of movie reviews.

\section{Dataset}

\begin{table}[]
\caption{The list of all data points considered for the ranking system}
\label{tab:dataset}
\resizebox{\textwidth}{!}{%
\begin{tabular}{|c|l|}
\hline
\textbf{}                    & Details about   the type of movie (Title, Overview, Genre, Tagline, Keywords) \\ \cline{2-2} 
\textbf{Metadata} &
  \begin{tabular}[c]{@{}l@{}}The factual   details of the movie (Runtime, Language, Director, \\ Cast, Writers, Production   Companies, Release Date, Budget)\end{tabular} \\ \cline{2-2} 
\textbf{} &
  \begin{tabular}[c]{@{}l@{}}The Response to   the movie from the audience (Popularity, Revenue, \\ Rating, Vote Count)\end{tabular} \\ \hline
\textbf{Visual   Similarity} & Official   Trailer for each movie                                             \\ \hline
\textbf{Sentiment Analysis}  & All   the user reviews for the given movie from IMDb                          \\ \hline
\end{tabular}%
}
\end{table}

One of the most exhaustive sources for metadata of any movie released worldwide is the Internet Movie Database (IMDb) \cite{imdb_ref}. To create a content-based recommendation system, we publicly make available a dataset of the top 10,000 English language-based movies sorted in descending order by their vote count till 13th August 2022. The metadata consists of all the data points mentioned in Table \ref{tab:dataset}. To rank the recommended movies based on their visual similarity, we utilise the official trailers of all the movies as extracted from YouTube \cite{youtube_ref}. In addition, for each of the recommended movies, all the user reviews from IMDb are considered for sentiment analysis. 

We tried experimenting with multilingual data by considering movies released in regional Indian languages. However, the feasibility of using such data drastically decreased because of its lack of uniformity and availability of credible sources. Nonetheless, we utilise information revolving around English-language-based movies only in the context of this paper.

\section{Methodology}
\begin{figure}[htp]
    \centering
    \includegraphics[width=9cm]{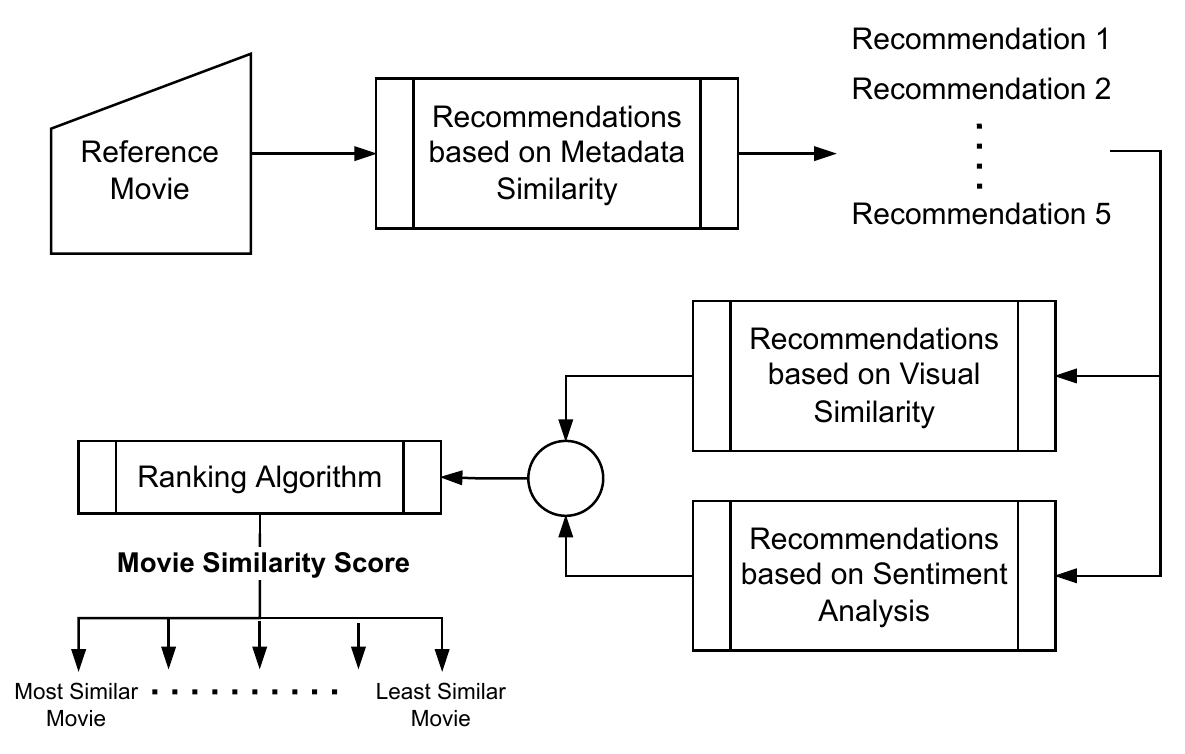}
    \caption{Flow diagram for the proposed recommendation system}
    \label{fig:overall_method}
\end{figure}

Fig \ref{fig:overall_method} refers to the flow diagram of our proposed system with three components, i.e., metadata similarity, visual similarity and sentiment analysis score.

\subsection{Metadata Similarity}
Before ranking the recommendations of a movie, we generate a list of recommendations for a random movie. We use a content-based filtering approach similar to Rujhan Singla et al. \cite{1-methodContent} in their paper. 

\subsubsection{Data Pre-Processing}

All of the columns are not required for generating recommendations, so a combination of the following metadata is vectorized and used -

\begin{equation}
    Combination = Keywords + Cast + Genres + Director + Overview
    \label{eq:combination_definition}
\end{equation}

In cases where we cannot get the required data (due to unavailability), we use empty strings to ensure continuity. Moreover, redundant information was removed by routine natural languages processing techniques such as removing stop-words and lemmatisation.

\subsubsection{Algorithm} 

The data used for this content-based recommendation system is in the form of text, but since machines cannot read strings, they require data to be numerical. To transform this raw textual data into a numerical format, we use a text vectorization algorithm, namely Term Frequency-Inverse Document Frequency or TF-IDF, introduced by Karen Sparck Jones \cite{2-methodContent}. The term frequency is the number of times a particular term appears in a document. The term frequency represents each text from the data as a matrix. The quantity of documents that use a particular term is known as document frequency. It indicates how common the term is. Inverse Document Frequency or IDF aims to reduce the weight of a term if it appears numerous times across all the documents. It can be calculated as follows:

\begin{equation}
    idf\textsubscript{i} = log(\frac{n}{df\textsubscript{i}})
\end{equation}

Here, idf\textsubscript{i} is the IDF score for term i, df\textsubscript{i} is the number of documents where the term i appears and n is the total number of documents under consideration. The TF-IDF score is the multiplication of the TF matrix with its IDF-

\begin{equation}
    w\textsubscript{i,j} = tf\textsubscript{i,j} \times idf\textsubscript{i}
\end{equation}

where W\textsubscript{i,j} is TF-IDF score for each term i in document j, tf\textsubscript{i,j} is term frequency for term i in document j, and idf\textsubscript{i} is IDF score for term i. 
The combination which is to be vectorized is shown in \eqref{eq:combination_definition}.

We create a single string for each movie consisting of the above features. These were chosen because they would provide the most value for deciding which movies are most similar to the given movie. In this scenario, a movie represents a document and the combination above denotes a term. In this case, the IDF of a document in the corpus will denote the number of documents or movies where words in the combination will appear. As explained in \cite{3-methodContent}, this will be used to assign less weight to terms that are used frequently. Cosine similarity is used to calculate the distance between the unit vectors of the movies. The movies having the shortest distance would be most similar to the initially given movie, as also used by D Gunawan et al. \cite{4-methodContent} to calculate the text relevance between two documents. Cosine similarity is the cosine of the angle between two vectors. 

\begin{equation}
    similarity = cos(\theta) = \frac{A.B}{|A| |B|} = \frac{\sum_{i=1}^{n} A\textsubscript{i}B\textsubscript{i}}  {\sqrt{\sum_{i=1}^{n} A\textsubscript{i}\textsuperscript{2}} \sqrt{\sum_{i=1}^{n} B\textsubscript{i}\textsuperscript{2}}}
\end{equation}
where A is the vector of the initially given movie and B is the vector of every other movie in the corpus.
We use the above-described content-based filter to further test our ranking algorithm to recommend five suggestions for three movies.

\subsection{Visual Similarity}
The primary approach to ranking a list of movies based on their visual similarity to the reference movie is based on a process analogous to that utilised by Yashar et al. \cite{20-lit} with a more unsupervised outlook rather than a supervised classification problem. Instead of extracting low-level stylistic features like colour, motion, and lighting, we rely on a clustering methodology to segment the movie's trailer.
There is also sufficient evidence implying that trailers and movies are highly correlated, allowing us to consider trailers to represent full-length movies. Hence we consider a movie's trailer to be a good indicator of the visual features in the movie. \\

\begin{figure}[htp]
    \centering
    \includegraphics[width=12cm]{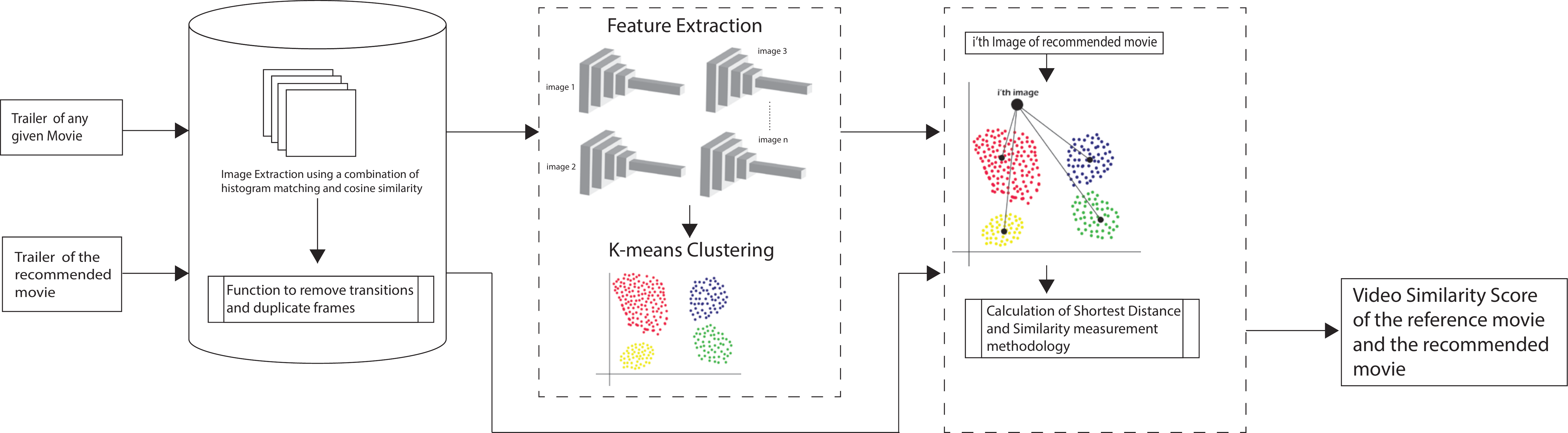}
    \caption{The flow diagram for the calculation of Video Similarity}
    \label{fig:video_similarity}
\end{figure}

\begin{algorithm}[H]
    \centering
    \caption{Calculation of Histogram difference for identifying Key Frames}
    \label{alg:algorithm1}
    \footnotesize
    \begin{algorithmic}[1]
        \State \text{Start Key Frame Extraction} 
        \State $REF \gets 1$
        \State $KeyFrameList \gets \{\}$
        \State $\textit{N} \gets \text{Total Frames}$

        \For{$i\gets 2, N$}
        
            \State $H1 \gets Histogram(Frame(REF))$
            \State $H2 \gets Histogram(Frame(i))$
            \If{$H1-H2 \geq \displaystyle\left\lvert 0.85 \right\rvert$ }
                \State  KeyFrameList.insert(Frame(i))
            \EndIf
            \State $REF \gets REF+1$
            
        \EndFor
        \State \textbf{Return}  \text{KeyFrameList}
    \end{algorithmic}
\end{algorithm}
\begin{algorithm}[H]
    \centering
    \caption{Calculation of Cosine Similarity For removing similar Keyframes}
    \label{alg:algorithm2}
    \footnotesize
    \begin{algorithmic}[1]
        \State \text{Start Key Frame Checking} 
        \State $REF \gets 1$
        \State $UniqueKeyFrames \gets \{\}$
        \State $\textit{N} \gets \text{Total Key Frames}$

        \For{$i\gets 2, N$}
        
            \State $C1 \gets Cosine(Frame(REF))$
            \State $C2 \gets Cosine(Frame(i))$
            \If{$CosineSimilarity(C1,C2) < \displaystyle\left\lvert 0.9 \right\rvert$ }
                \State  UniqueKeyFrames.insert(Frame(i))
            \State $REF \gets REF+1$
            \EndIf
            
        \EndFor
        \State \textbf{Return}  \text{UniqueKeyFrames}
    \end{algorithmic}
\end{algorithm}

\subsubsection{Image Extraction}
The extraction of frames is based on a key frame extraction model \cite{1-methodVideo} except that instead of only histogram matching, we implement a combination of histogram matching and a cosine similarity metric. This approach enables an understanding of whether the frames have sufficient new information compared to the previous frame and ensures that similar key frames are not extracted. Moreover, since more recent movies are shot at higher frames per second, extracting all frames would create a high correlation and be computationally expensive.
Though there are many ways of extracting frames from a video, we use a fixed interval extraction before proceeding with the key frame detection. We extract frames alternatingly as mentioned in Algorithms \ref{alg:algorithm1} and \ref{alg:algorithm2}. Based on a preliminary analysis of 100 randomly selected trailers from the dataset, the average duration is 120–180 seconds and the average frame rate is 30 frames per second. As a result, the total number of extracted frames (without additional filters) is between 2000 and 2500.


\subsubsection{Data Preprocessing}
Good movie trailers are designed to pique the viewer's curiosity and show the quality of the movie \cite{4-methodVideo}. For the same reasons, production houses often keep many cut scenes, transitions, introductory animation, ending animation, and text-based frame animation during cut scenes. However, the visual cues provided in these frames are negligible as the relevant information is already part of the metadata. Additionally, as these frames can interfere with the algorithm, we implement filtering criteria to eliminate such frames. For transitions and introductory animation, we utilise PySceneDetect \cite{5-methodVideo}; for images that are pure color (cut-scenes, transitions), we empirically determine the threshold of the pixel intensities and the number of such pixels that do not provide any extra information about the trailer. We tested this approach manually on 20 trailers to determine the thresholds. Each black and white frame is composed of a set number of pixels having intensity between 1 to 255, with the former being a pitch dark pixel and the latter being pure white pixel. The result is shown in Table \ref{tab:pixel}.

\begin{table}[]
\caption{The absolute pixel intensities for black images and white images}
\label{tab:pixel}
\resizebox{\textwidth}{!}{%
\begin{tabular}{|cc|lcc}
\cline{1-2} \cline{4-5}
\multicolumn{2}{|c|}{\textbf{Mostly Black: 16 samples}} & \multicolumn{1}{l|}{} & \multicolumn{2}{c|}{\textbf{Mostly White: 11   samples}}        \\ \cline{1-2} \cline{4-5} 
\multicolumn{1}{|c|}{\textbf{Pixel Intensity}} &
  \textbf{\begin{tabular}[c]{@{}c@{}}Number of Images \\ with 80\% pixels under \\ the intensity\end{tabular}} &
  \multicolumn{1}{l|}{} &
  \multicolumn{1}{c|}{\textbf{Pixel Intensity}} &
  \multicolumn{1}{c|}{\textbf{\begin{tabular}[c]{@{}c@{}}Number of Images \\ with 80\% pixels under \\ the intensity\end{tabular}}} \\ \cline{1-2} \cline{4-5} 
\multicolumn{1}{|c|}{\textbf{1}}            & 5         & \multicolumn{1}{l|}{} & \multicolumn{1}{c|}{\textbf{215}}     & \multicolumn{1}{c|}{2}  \\ \cline{1-2} \cline{4-5} 
\multicolumn{1}{|c|}{\textbf{2}}            & 6         & \multicolumn{1}{l|}{} & \multicolumn{1}{c|}{\textbf{216-223}} & \multicolumn{1}{c|}{5}  \\ \cline{1-2} \cline{4-5} 
\multicolumn{1}{|c|}{\textbf{3-9}}          & 8         & \multicolumn{1}{l|}{} & \multicolumn{1}{c|}{\textbf{224-235}} & \multicolumn{1}{c|}{6}  \\ \cline{1-2} \cline{4-5} 
\multicolumn{1}{|c|}{\textbf{10-12}}        & 9         & \multicolumn{1}{l|}{} & \multicolumn{1}{c|}{\textbf{236-241}} & \multicolumn{1}{c|}{7}  \\ \cline{1-2} \cline{4-5} 
\multicolumn{1}{|c|}{\textbf{13-16}}        & 10        & \multicolumn{1}{l|}{} & \multicolumn{1}{c|}{\textbf{242-253}} & \multicolumn{1}{c|}{8}  \\ \cline{1-2} \cline{4-5} 
\multicolumn{1}{|c|}{\textbf{17}}           & 11        & \multicolumn{1}{l|}{} & \multicolumn{1}{c|}{\textbf{254}}     & \multicolumn{1}{c|}{9}  \\ \cline{1-2} \cline{4-5} 
\multicolumn{1}{|c|}{\textbf{18}}           & 12        & \multicolumn{1}{l|}{} & \multicolumn{1}{c|}{\textbf{255}}     & \multicolumn{1}{c|}{11} \\ \cline{1-2} \cline{4-5} 
\multicolumn{1}{|c|}{\textbf{19-21}}        & 13        &                       & \multicolumn{1}{l}{}                  & \multicolumn{1}{l}{}    \\ \cline{1-2}
\multicolumn{1}{|c|}{\textbf{22-26}}        & 14        &                       & \multicolumn{1}{l}{}                  & \multicolumn{1}{l}{}    \\ \cline{1-2}
\multicolumn{1}{|c|}{\textbf{27-30}}        & 15        &                       & \multicolumn{1}{l}{}                  & \multicolumn{1}{l}{}    \\ \cline{1-2}
\multicolumn{1}{|c|}{\textbf{31-33}}        & 16        &                       & \multicolumn{1}{l}{}                  & \multicolumn{1}{l}{}    \\ \cline{1-2}
\end{tabular}%
}
\end{table}

We use a Laplacian Operator for blurred images to determine edges within the picture \cite{9-methodVideo}. Based on multiple iterations, the threshold we chose is 2, i.e., we filter out all those frames that have a variance of Laplacian less than 2
Additionally, not all movies are shot in the same camera setting. Due to different aspect ratios and perceptive cinematography, the presence of letterboxes and cinematic black bars causes a difference in the frames of different movies. Our framework takes care of this using image-processing functions provided by the OpenCV library.

\subsubsection{Feature Extraction}


\begin{figure}[htp]
    \centering
    \includegraphics[width=7cm]{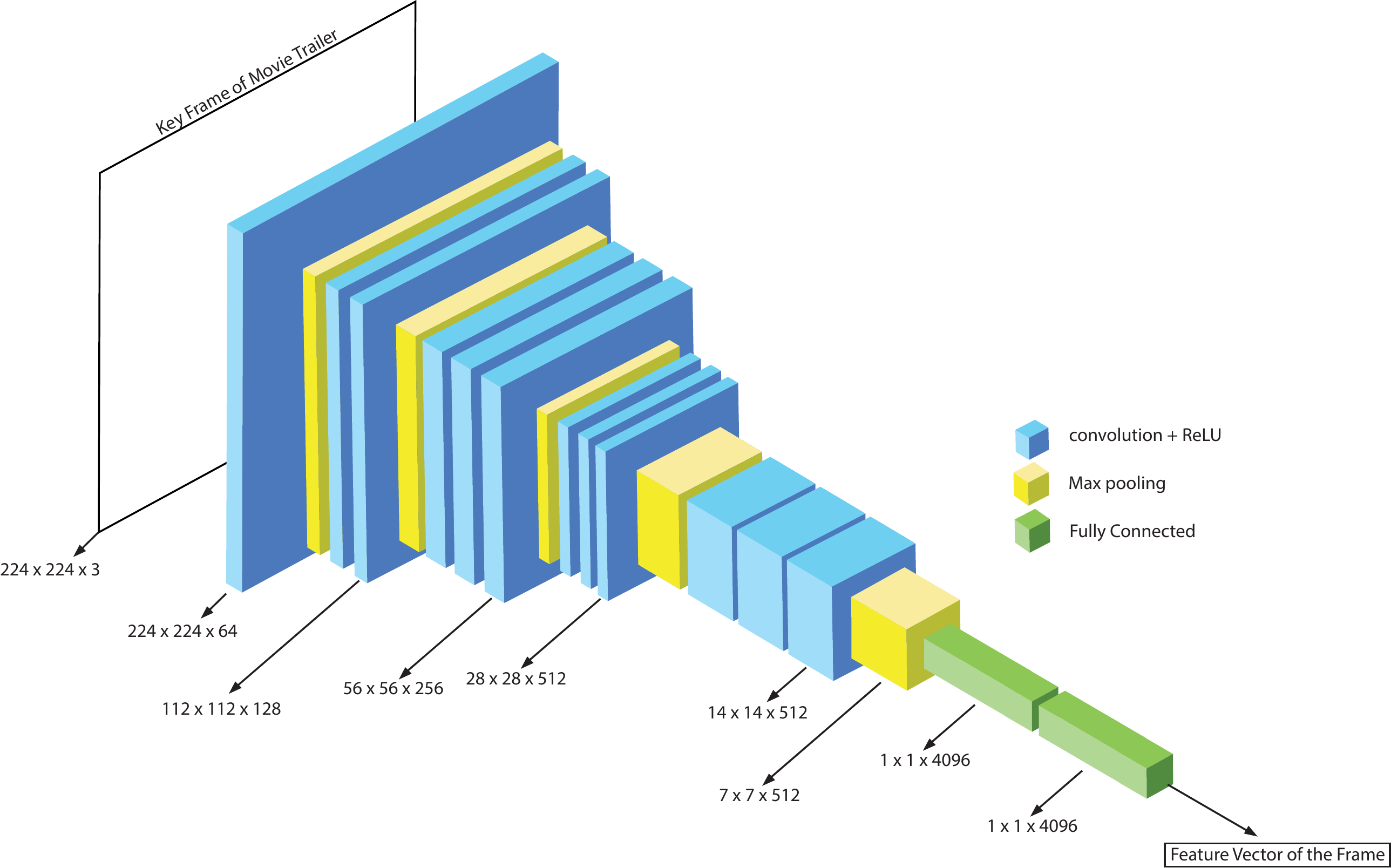}
    \caption{Breakdown of the VGG Convolutional Network used for feature extraction}
    \label{fig:vgg_diagram}
\end{figure}

We use a pre-trained Visual Geometry Group (VGG) network architecture \cite{6-methodVideo} for extracting the most important stylistic features from the database of keyframes of the reference trailer. It is based on a CNN model, trained on the magnanimous ImageNet dataset, and performs considerably well in creating feature vectors for a given image \cite{8-methodVideo}. VGG19 inputs all images with a dimension of 224x224x3 by default. The architecture generates a feature vector with dimensions of 1x1x4096 for each frame (see
Fig.~\ref{fig:vgg_diagram}). Key frames within movie trailers store detailed information, and resizing them to a size of 224x224x3 before extracting features can cause some of these features to lose information. However, the tradeoff is that generating features from high-resolution images is computationally expensive.

\subsubsection{Kmeans Clustering}
After extracting the feature vectors of multiple key frames, we use a K-means clustering algorithm to cluster all the feature vectors \cite{10-methodVideo}. This step helps segment the type of scenes in the trailer (each key frame represents a scene).
We try to select a fixed number of clusters based on the Elbow method \cite{11-methodVideo}. However, due to the fast pace of movie trailers and significant differences between most of the scenes, most of them start showing minimum deviation in WCSS as the number of clusters approaches the total length of extracted key frames. Hence, we fix the number of clusters at 5, where we manually gauge the clusters to be significantly different.

\subsubsection{Calculation of Similarity}
After the query frame is preprocessed, we find the euclidean distance between the feature vector of the query frame and the centroid of all the clusters of the reference movie. The cluster centroid that is closest to the feature vector of the query frame exhibits the highest similarity compared to other centroids. Repeating this approach for all the frames of the query trailer, we get a value of a cluster centroid of the reference movie for each frame of the query trailer. Now we compare the percentage of query frames in each cluster compared to the total frames in the query trailer. By doing this, we get a distribution vector as shown in Equation \ref{eq:point}. With this information, we compare the distribution vector of the query trailer with that of the reference trailer by using Equation \ref{eq:euclidean_distance_for_vss}.

\begin{equation}
 P= (v_{1},w_{1},x_{1},y_{1},z_{1})
 \label{eq:point}
\end{equation}      
where,

\begin{center}
    v\textsubscript{1}, w\textsubscript{1},.....,z\textsubscript{1} signify the percentage distribution of key frames into clusters
\end{center}

\begin{equation}
   Euclidean \ Distance = \sqrt {\left( {v_1 - v_2 } \right)^2 + \left( {w_1 - w_2 } \right)^2 ........ \left( {z_1 - z_2 } \right)^2 }
   \label{eq:euclidean_distance_for_vss}
\end{equation}

\begin{figure}[htp]
    \centering
    \includegraphics[width=11cm]{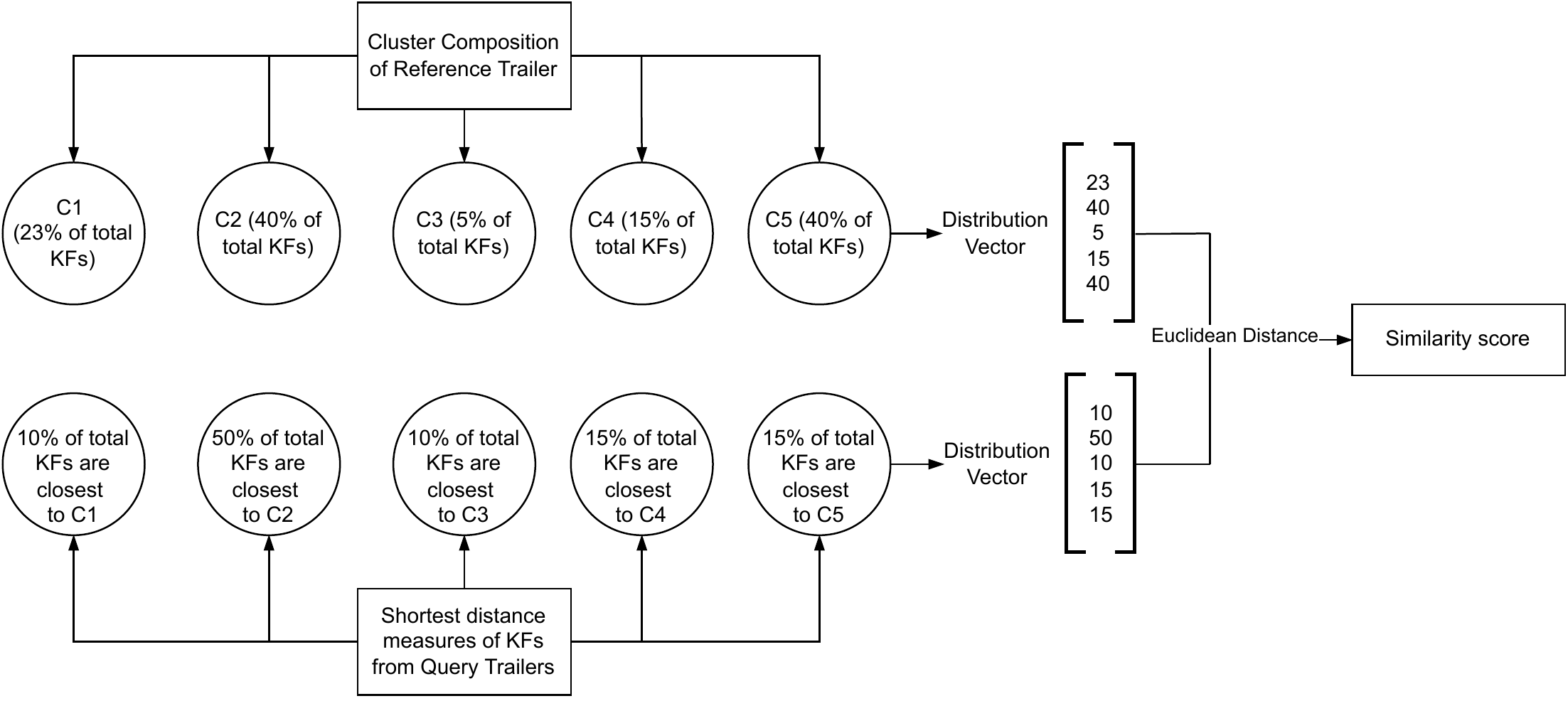}
    \caption{The calculation of Visual Similarity}
    \label{fig:calculation_of_video_similarity}
\end{figure}

Based on the Euclidean distance between the two distributions, we quantify how close the two points are. The inverse of this metric is what we call the visual similarity score as mentioned in Equation \ref{eq:vss_calculation}. The complete calculation methodology is visualised in
Fig.~\ref{fig:calculation_of_video_similarity}.

\begin{equation}
        VSS = \frac{1}{1 + x}   \\  
        \label{eq:vss_calculation}
\end{equation}
where,

\begin{center}
    x is Euclidean distance
\end{center}

\subsection{Sentiment Analysis}
A movie review is a piece of writing that expresses the writers' opinions about a particular film and offers either support or criticism, allowing a viewer to decide whether or not they want to see the movie. Such reviews act as indirect recommendations to the viewer. To better collect, retrieve, measure, and evaluate viewers, it is crucial to be able to classify movie reviews \cite{1-methodSentiment}. Thus, we define sentiment analysis of movie reviews as a classification problem and attempt to solve it as presented by Mais Yasen et al. \cite{7-methodSentiment}.  

\subsubsection{Data Pre-Processing}

In order to prepare data for training a machine learning model, we need to process it by removing all HTML tags, punctuations, single characters and multiple spaces.

Since a machine learning algorithm cannot process direct text, we use Count Vectorizer to convert the reviews into vectors.

\subsubsection{Algorithms Used}

After obtaining the text in vector form, we use TF-IDF to get the importance of each word in the review. The data is split into train and test, and we test several classification algorithms to see which gives us the best result. The algorithms used are-

\paragraph{\textbf{Linear Support Vector Classifier (Linear SVC)}}
This algorithm classifies data using a linear kernel function and performs well when many samples are involved. This algorithm aims to find a hyperplane that will separate the given samples into two classes in a P-dimension space.


\paragraph{\textbf{Logistic Regression}}
Introduced in 1958 \cite{2-methodSentiment}, it is one of the earliest methods invented to perform classification. Logistic regression measures the relationship between the categorical dependent variable and one or more independent variables by estimating probabilities using a sigmoid curve.

\begin{equation}
        f(x) = \frac{L}{1 + e \textsuperscript{-k(x-x\textsubscript{0})}}   \\       
\end{equation}

\begin{center}
    \text{Sigmoid Curve Equation}
\end{center}

\paragraph{\textbf{Decision Trees}}
This algorithm uses a tree-like graph or model of decisions and their possible consequences. It has a
flow-like structure in which each internal node represents a "test" on an attribute \cite{3-methodSentiment}.

\begin{equation}
    Entropy(S) = - \sum_{}{}P(I) \times log\textsubscript{2}(P(I)) 
\end{equation}

\begin{equation}
    Information \ Gain(S,A) = Entropy(S) - \sum_{}{}P(S|A) \times Entropy(S|A)
\end{equation}

Entropy is the quantity of data required to describe a sample accurately. Information gain is the amount of information provided by a particular feature.

\paragraph{\textbf{Random Forest Classifier}}
This algorithm is an extension of the Decision Tree algorithm \cite{4-methodSentiment}. It works by creating more than one decision tree, which is equivalent to a ‘forest’. Like more trees in the forest, the more robust the forest looks, the higher the number of decision trees in a forest and the higher the accuracy.


\paragraph{\textbf{XGBoost}}
This algorithm is based on the decision tree algorithm that uses the gradient boosting framework \cite{5-methodSentiment}. Decision trees are created in a sequential form. Each independent variable is weighed before being fed into the decision tree that forecasts outcomes.  

\paragraph{\textbf{Naive Bayes}}
It is a group of linear classifiers that are simple and efficient.
The probabilistic model of this algorithm is based on the
Bayes Theorem \cite{6-methodSentiment}, and the adjective model comes from the
the fact that the features in a dataset are mutually
independent. 
    
\begin{equation}
    P(A|B) = \frac{P(B|A) P(A)}{P(B)}
\end{equation}
\begin{center}
    \text{Bayes Theorem}
\end{center}

\subsubsection{Metrics}

We use two evaluation metrics, accuracy and F1 score.

Accuracy : Ratio of correctly predicted observations to the total observations.

F1 Score : Weighted average of Precision and Recall.

\emph{Note: Here precision is the ratio of correctly predicted positive
observations to the total predicted positive observations, and
recall is the ratio of correctly predicted positive observations to
the total predicted positive observations.}

\begin{equation}
    Precision = \frac{TP}{TP + FP}
\end{equation}
\begin{equation}
    Recall = \frac{TP}{TP + FN}
\end{equation}
\begin{equation}
     F1 \ Score = \frac{2 \times Precision \times Recall}{Precision + Recall}
\end{equation}
\begin{equation}
    Accuracy = \frac{TP + TN}{TP + FN + TN + FP}
\end{equation}

\emph{where TP, TN, FP, FN are True Positive, True Negative, True Negative and  False Negative respectively.}


After finding out the best-performing algorithm, we use that algorithm to do sentiment analysis on the movies that we obtain from the content-based filtering approach discussed previously. Since the number of reviews available for each movie will not be the same (lower ranked movies have fewer reviews), we randomly sample 50 reviews at one time for each movie and repeat this process 10 times to find the percentage of positive and negative reviews. This ensures that the testing is unbiased due to the fluctuations in the number of reviews across different recommendations. We calculate the positivity score based on the average percentage of positive reviews of those 10 runs to determine the overall sentiment of a movie. The sentiment classification algorithm has been trained on a publicly available IMDb dataset \cite{imdbDS}.

\subsection{Calculation of Movie Similarity Score}\label{ssec:movie_similarity_calculation}

To generate a ranking of the movies recommended by our system, we have created a metric that consists of a weighted sum of the sentiment score and the visual similarity score. To decide the weights, we assume that the sentiment of the audience is unrelated to the visual similarity of the content. Hence we keep the weights to be 0.5 each giving equal importance to visual similarity and the sentiment analysis of the movies. For visual similarity, we consider the similarity between trailers of the reference and recommended movies based on the methodology proposed above. We have considered the percentage of positive reviews for a movie as the sentiment analysis score. 

\begin{equation}
    Movie \ Similarity \ Score = (VSS \times 0.5) + (SC \times 0.5)
\end{equation}
where,

\begin{center}
    VSS is Visual Similarity Score and SC is the Sentiment Score
\end{center}

\section{Results}
We have taken 3 random movies  \cite{tenet,castaway,2001spaceodyssey} from the dataset and generated 5 movies as recommendations for each of those movies \cite{interstellar,the355,predestination,manfromuncle,mifallout,6days7nights,lordoftheflies,nimsisland,bluelagoon,mostdangerousgame,darkstar,2010yearwemakecontact,gravity,blackhole,adastra}  as listed in Table \ref{tab:metadata_predictions}.

\begin{table}[]
\caption{The list of recommendations of the reference movies based on metadata similarity}
\label{tab:metadata_predictions}
\begin{tabular}{l|ccccc|}
\cline{2-6}
\textbf{} &
  \multicolumn{5}{c|}{\textbf{Recommended Movies}} \\ \hline
\multicolumn{1}{|l|}{\textbf{Reference Movies}} &
  \multicolumn{1}{c|}{\textbf{1}} &
  \multicolumn{1}{c|}{\textbf{2}} &
  \multicolumn{1}{c|}{\textbf{3}} &
  \multicolumn{1}{c|}{\textbf{4}} &
  \textbf{5} \\ \hline
\multicolumn{1}{|c|}{\textbf{Tenet}} &
  \multicolumn{1}{c|}{Interstellar} &
  \multicolumn{1}{c|}{\begin{tabular}[c]{@{}c@{}}The\\ 355\end{tabular}} &
  \multicolumn{1}{c|}{Predestination} &
  \multicolumn{1}{c|}{\begin{tabular}[c]{@{}c@{}}The   Man \\ from \\ U.N.C.L.E\end{tabular}} &
  \begin{tabular}[c]{@{}c@{}}Mission   \\ Impossible: \\ Fallout\end{tabular} \\ \hline
\multicolumn{1}{|c|}{\textbf{Cast Away}} &
  \multicolumn{1}{c|}{\begin{tabular}[c]{@{}c@{}}Six   Days \\ Seven \\ Nights\end{tabular}} &
  \multicolumn{1}{c|}{\begin{tabular}[c]{@{}c@{}}Lord of \\ the Flies\end{tabular}} &
  \multicolumn{1}{c|}{Nim's   Island} &
  \multicolumn{1}{c|}{\begin{tabular}[c]{@{}c@{}}The   Blue \\ Lagoon\end{tabular}} &
  \begin{tabular}[c]{@{}c@{}}The   Most \\ Dangerous\\ Game\end{tabular} \\ \hline
\multicolumn{1}{|c|}{\textbf{\begin{tabular}[c]{@{}l@{}}2001: A Space   \\ Odyssey\end{tabular}}} &
  \multicolumn{1}{c|}{\begin{tabular}[c]{@{}c@{}}Dark\\ Star\end{tabular}} &
  \multicolumn{1}{c|}{\begin{tabular}[c]{@{}c@{}}2010:   The \\ Year We Make \\ Contact\end{tabular}} &
  \multicolumn{1}{c|}{Gravity} &
  \multicolumn{1}{c|}{\begin{tabular}[c]{@{}c@{}}The   Black \\ Hole\end{tabular}} &
  Ad   Astra \\ \hline
\end{tabular}
\end{table}

\subsection{Visual Similarity Score}
For all the 18 movie titles, we compared the trailer of the recommended movies to the reference movies. The average length of each trailer is 130 seconds and the average length of key frames is 210. 
There is a reduction of 20\% of the key frames after using a cosine similarity approach to remove matching frames. The final distribution metric gives the following video similarity score for all the recommended movie trailers compared to the reference movie trailer.

\begin{table}[]
\caption{The Visual Similarity score of all recommended movies}
\label{tab:my-table}
\resizebox{\textwidth}{!}{%
\begin{tabular}{|lclclc|}
\hline
\multicolumn{6}{|c|}{\textbf{Video Similarity of Recommended Movies}} \\ \hline
\multicolumn{2}{|c|}{\textbf{Tenet}} &
  \multicolumn{2}{c|}{\textbf{CastAway}} &
  \multicolumn{2}{c|}{\textbf{2001: Space Odyssey}} \\ \hline
\multicolumn{1}{|l|}{Mission Impossible: Fallout} &
  \multicolumn{1}{c|}{\textbf{0.933}} &
  \multicolumn{1}{l|}{Nims Island} &
  \multicolumn{1}{c|}{\textbf{0.871}} &
  \multicolumn{1}{l|}{Gravity} &
  \textbf{0.803} \\ \hline
\multicolumn{1}{|l|}{Predestination} &
  \multicolumn{1}{c|}{\textbf{0.892}} &
  \multicolumn{1}{l|}{Most Dangerous Game} &
  \multicolumn{1}{c|}{\textbf{0.805}} &
  \multicolumn{1}{l|}{Ad Astra} &
  \textbf{0.771} \\ \hline
\multicolumn{1}{|l|}{The Man from U.N.C.L.E} &
  \multicolumn{1}{c|}{\textbf{0.879}} &
  \multicolumn{1}{l|}{Six Days Seven Nights} &
  \multicolumn{1}{c|}{\textbf{0.782}} &
  \multicolumn{1}{l|}{Dark Star} &
  \textbf{0.739} \\ \hline
\multicolumn{1}{|l|}{The 355} &
  \multicolumn{1}{c|}{\textbf{0.825}} &
  \multicolumn{1}{l|}{Lord of The Flies} &
  \multicolumn{1}{c|}{\textbf{0.764}} &
  \multicolumn{1}{l|}{Black Hole} &
  \textbf{0.733} \\ \hline
\multicolumn{1}{|l|}{Interstellar} &
  \multicolumn{1}{c|}{\textbf{0.813}} &
  \multicolumn{1}{l|}{Blue Lagoon} &
  \multicolumn{1}{c|}{\textbf{0.754}} &
  \multicolumn{1}{l|}{Contact 2010} &
  \textbf{0.686} \\ \hline
\end{tabular}%
}
\end{table}

\subsection{Sentiment Score}

Table \ref{tab:model_performance} shows a comparative study for multiple sentiment classification algorithms. Table \ref{tab:tenet_sentiment}, \ref{tab:cast_away_sentiment} and \ref{tab:2001_space_sentiment}, show a summary of how we achieved a sentiment score for each movie.
\begin{table}[]
\caption{Comparative study of the performance of different machine learning models on the IMDb dataset}
\label{tab:model_performance}
\resizebox{\textwidth}{!}{%
\begin{tabular}{|c|l|c|c|c|c|}
\hline
\multicolumn{1}{|l|}{\textbf{Sr. No.}} &
  \textbf{Algorithm} &
  \multicolumn{1}{l|}{\textbf{Accuracy}} &
  \multicolumn{1}{l|}{\textbf{Precision}} &
  \multicolumn{1}{l|}{\textbf{Recall}} &
  \multicolumn{1}{l|}{\textbf{F1-Score}} \\ \hline
1 & \textbf{Linear Support Vector Classification} & 90\%    & 90.79\% & 88.92\% & 89.85\% \\ \hline
2 & \textbf{Logistic Regression}                  & 89.66\% & 83.86\% & 87.52\% & 85.65\% \\ \hline
3 & \textbf{Decision Tree}                        & 71.52\% & 71.40\% & 71.33\% & 71.36\% \\ \hline
4 & \textbf{Random Forest Classifier}             & 74.58\% & 70.77\% & 83.06\% & 76.43\% \\ \hline
5 & \textbf{XGBoost}                              & 86.07\% & 87.51\% & 83.97\% & 85.71\% \\ \hline
6 & \textbf{Naive Bayes}                          & 85.41\% & 83.86\% & 87.52\% & 85.65\% \\ \hline
\end{tabular}%
}
\end{table}

Based on table \ref{tab:model_performance}, Linear Support Vector Classifier gives the best result on the IMDb data with an accuracy of 90\% and F1 Score of 89.85\%.

 
\begin{table}[]
\caption{The sentiment-level ranking of recommended movies for Tenet}
\label{tab:tenet_sentiment}
\resizebox{\textwidth}{!}{%
\begin{tabular}{l|cc|cc|cc|cc|cc|lll}
\cline{2-11}
 &
  \multicolumn{2}{c|}{\textbf{Interstellar}} &
  \multicolumn{2}{c|}{\textbf{The   355}} &
  \multicolumn{2}{c|}{\textbf{Predestination}} &
  \multicolumn{2}{c|}{\textbf{\begin{tabular}[c]{@{}c@{}}The   Man From \\ U.N.C.L.E\end{tabular}}} &
  \multicolumn{2}{c|}{\textbf{\begin{tabular}[c]{@{}c@{}}Mission   Impossible: \\ Fallout\end{tabular}}} &
  &
  &
  \\ \cline{2-11} \cline{13-14} 
 &
  \multicolumn{1}{c|}{\textbf{Positive}} &
  \textbf{Negative} &
  \multicolumn{1}{c|}{\textbf{Positive}} &
  \textbf{Negative} &
  \multicolumn{1}{c|}{\textbf{Positive}} &
  \textbf{Negative} &
  \multicolumn{1}{c|}{\textbf{Positive}} &
  \textbf{Negative} &
  \multicolumn{1}{c|}{\textbf{Positive}} &
  \textbf{Negative} &
  \multicolumn{1}{l|}{} &
  \multicolumn{2}{c|}{\textbf{Ranking}} \\ \cline{1-11} \cline{13-14} 
\multicolumn{1}{|l|}{\textbf{Run 1}} &
  \multicolumn{1}{c|}{72\%} &
  28\% &
  \multicolumn{1}{c|}{56\%} &
  44\% &
  \multicolumn{1}{c|}{72\%} &
  28\% &
  \multicolumn{1}{c|}{64\%} &
  36\% &
  \multicolumn{1}{c|}{70\%} &
  30\% &
  \multicolumn{1}{l|}{} &
  \multicolumn{1}{c|}{\textbf{1}} &
  \multicolumn{1}{l|}{Predestination} \\ \cline{1-11} \cline{13-14} 
\multicolumn{1}{|l|}{\textbf{Run 2}} &
  \multicolumn{1}{c|}{72\%} &
  28\% &
  \multicolumn{1}{c|}{40\%} &
  60\% &
  \multicolumn{1}{c|}{78\%} &
  22\% &
  \multicolumn{1}{c|}{68\%} &
  32\% &
  \multicolumn{1}{c|}{68\%} &
  32\% &
  \multicolumn{1}{l|}{} &
  \multicolumn{1}{c|}{\textbf{2}} &
  \multicolumn{1}{l|}{Interstellar} \\ \cline{1-11} \cline{13-14} 
\multicolumn{1}{|l|}{\textbf{Run 3}} &
  \multicolumn{1}{c|}{74\%} &
  26\% &
  \multicolumn{1}{c|}{46\%} &
  54\% &
  \multicolumn{1}{c|}{68\%} &
  32\% &
  \multicolumn{1}{c|}{60\%} &
  40\% &
  \multicolumn{1}{c|}{64\%} &
  36\% &
  \multicolumn{1}{l|}{} &
  \multicolumn{1}{c|}{\textbf{3}} &
  \multicolumn{1}{l|}{Mission Impossible: Fallout} \\ \cline{1-11} \cline{13-14} 
\multicolumn{1}{|l|}{\textbf{Run 4}} &
  \multicolumn{1}{c|}{70\%} &
  30\% &
  \multicolumn{1}{c|}{46\%} &
  54\% &
  \multicolumn{1}{c|}{82\%} &
  18\% &
  \multicolumn{1}{c|}{66\%} &
  34\% &
  \multicolumn{1}{c|}{54\%} &
  46\% &
  \multicolumn{1}{l|}{} &
  \multicolumn{1}{c|}{\textbf{4}} &
  \multicolumn{1}{l|}{The Man From U.N.C.L.E} \\ \cline{1-11} \cline{13-14} 
\multicolumn{1}{|l|}{\textbf{Run 5}} &
  \multicolumn{1}{c|}{62\%} &
  38\% &
  \multicolumn{1}{c|}{48\%} &
  52\% &
  \multicolumn{1}{c|}{64\%} &
  36\% &
  \multicolumn{1}{c|}{60\%} &
  40\% &
  \multicolumn{1}{c|}{72\%} &
  28\% &
  \multicolumn{1}{l|}{} &
  \multicolumn{1}{c|}{\textbf{5}} &
  \multicolumn{1}{l|}{The 355} \\ \cline{1-11} \cline{13-14} 
\multicolumn{1}{|l|}{\textbf{Run 6}} &
  \multicolumn{1}{c|}{68\%} &
  32\% &
  \multicolumn{1}{c|}{42\%} &
  58\% &
  \multicolumn{1}{c|}{74\%} &
  26\% &
  \multicolumn{1}{c|}{74\%} &
  26\% &
  \multicolumn{1}{c|}{68\%} &
  32\% &
  &
  &
  \\ \cline{1-11}
\multicolumn{1}{|l|}{\textbf{Run 7}} &
  \multicolumn{1}{c|}{60\%} &
  40\% &
  \multicolumn{1}{c|}{46\%} &
  54\% &
  \multicolumn{1}{c|}{72\%} &
  28\% &
  \multicolumn{1}{c|}{52\%} &
  48\% &
  \multicolumn{1}{c|}{74\%} &
  26\% &
  &
  &
  \\ \cline{1-11}
\multicolumn{1}{|l|}{\textbf{Run 8}} &
  \multicolumn{1}{c|}{54\%} &
  46\% &
  \multicolumn{1}{c|}{36\%} &
  64\% &
  \multicolumn{1}{c|}{70\%} &
  30\% &
  \multicolumn{1}{c|}{58\%} &
  42\% &
  \multicolumn{1}{c|}{50\%} &
  50\% &
  &
  &
  \\ \cline{1-11}
\multicolumn{1}{|l|}{\textbf{Run 9}} &
  \multicolumn{1}{c|}{70\%} &
  30\% &
  \multicolumn{1}{c|}{40\%} &
  60\% &
  \multicolumn{1}{c|}{74\%} &
  26\% &
  \multicolumn{1}{c|}{74\%} &
  26\% &
  \multicolumn{1}{c|}{60\%} &
  40\% &
  &
  &
  \\ \cline{1-11}
\multicolumn{1}{|l|}{\textbf{Run 10}} &
  \multicolumn{1}{c|}{66\%} &
  34\% &
  \multicolumn{1}{c|}{52\%} &
  48\% &
  \multicolumn{1}{c|}{72\%} &
  28\% &
  \multicolumn{1}{c|}{64\%} &
  36\% &
  \multicolumn{1}{c|}{64\%} &
  36\% &
  &
  &
  \\ \cline{1-11}
\multicolumn{1}{|l|}{\textbf{Average}} &
  \multicolumn{1}{c|}{\textbf{67\%}} &
  33\% &
  \multicolumn{1}{c|}{\textbf{45\%}} &
  55\% &
  \multicolumn{1}{c|}{\textbf{73\%}} &
  27\% &
  \multicolumn{1}{c|}{\textbf{64\%}} &
  36\% &
  \multicolumn{1}{c|}{\textbf{64\%}} &
  36\% &
  &
  &
  \\ \cline{1-11}
\end{tabular}%
}
\end{table}

\begin{table}[]
\caption{The sentiment-level ranking of recommended movies for Cast Away}
\label{tab:cast_away_sentiment}
\resizebox{\textwidth}{!}{%
\begin{tabular}{l|cc|cc|cc|cc|cc|lll}
\cline{2-11}
 &
  \multicolumn{2}{c|}{\textbf{Six Days Seven Nights}} &
  \multicolumn{2}{c|}{\textbf{Lord   of the Flies}} &
  \multicolumn{2}{c|}{\textbf{Nim's   Island}} &
  \multicolumn{2}{c|}{\textbf{The   Blue Lagoon}} &
  \multicolumn{2}{c|}{\textbf{\begin{tabular}[c]{@{}c@{}}The   Most Dangerous \\ Game\end{tabular}}} &
  &
  &
  \\ \cline{2-11} \cline{13-14} 
 &
  \multicolumn{1}{c|}{\textbf{Positive}} &
  \textbf{Negative} &
  \multicolumn{1}{c|}{\textbf{Positive}} &
  \textbf{Negative} &
  \multicolumn{1}{c|}{\textbf{Positive}} &
  \textbf{Negative} &
  \multicolumn{1}{c|}{\textbf{Positive}} &
  \textbf{Negative} &
  \multicolumn{1}{c|}{\textbf{Positive}} &
  \textbf{Negative} &
  \multicolumn{1}{l|}{} &
  \multicolumn{2}{c|}{\textbf{Ranking}} \\ \cline{1-11} \cline{13-14} 
\multicolumn{1}{|l|}{\textbf{Run 1}} &
  \multicolumn{1}{c|}{46\%} &
  54\% &
  \multicolumn{1}{c|}{52\%} &
  48\% &
  \multicolumn{1}{c|}{74\%} &
  26\% &
  \multicolumn{1}{c|}{70\%} &
  30\% &
  \multicolumn{1}{c|}{70\%} &
  30\% &
  \multicolumn{1}{l|}{} &
  \multicolumn{1}{c|}{\textbf{1}} &
  \multicolumn{1}{l|}{Nim's Island} \\ \cline{1-11} \cline{13-14} 
\multicolumn{1}{|l|}{\textbf{Run 2}} &
  \multicolumn{1}{c|}{46\%} &
  54\% &
  \multicolumn{1}{c|}{44\%} &
  56\% &
  \multicolumn{1}{c|}{74\%} &
  26\% &
  \multicolumn{1}{c|}{78\%} &
  22\% &
  \multicolumn{1}{c|}{84\%} &
  16\% &
  \multicolumn{1}{l|}{} &
  \multicolumn{1}{c|}{\textbf{2}} &
  \multicolumn{1}{l|}{The Most Dangerous   Game} \\ \cline{1-11} \cline{13-14} 
\multicolumn{1}{|l|}{\textbf{Run 3}} &
  \multicolumn{1}{c|}{58\%} &
  42\% &
  \multicolumn{1}{c|}{48\%} &
  52\% &
  \multicolumn{1}{c|}{74\%} &
  26\% &
  \multicolumn{1}{c|}{64\%} &
  36\% &
  \multicolumn{1}{c|}{76\%} &
  24\% &
  \multicolumn{1}{l|}{} &
  \multicolumn{1}{c|}{\textbf{3}} &
  \multicolumn{1}{l|}{The Blue Lagoon} \\ \cline{1-11} \cline{13-14} 
\multicolumn{1}{|l|}{\textbf{Run 4}} &
  \multicolumn{1}{c|}{52\%} &
  48\% &
  \multicolumn{1}{c|}{38\%} &
  62\% &
  \multicolumn{1}{c|}{70\%} &
  30\% &
  \multicolumn{1}{c|}{72\%} &
  28\% &
  \multicolumn{1}{c|}{80\%} &
  20\% &
  \multicolumn{1}{l|}{} &
  \multicolumn{1}{c|}{\textbf{4}} &
  \multicolumn{1}{l|}{Six Days Seven Nights} \\ \cline{1-11} \cline{13-14} 
\multicolumn{1}{|l|}{\textbf{Run 5}} &
  \multicolumn{1}{c|}{50\%} &
  50\% &
  \multicolumn{1}{c|}{46\%} &
  54\% &
  \multicolumn{1}{c|}{82\%} &
  18\% &
  \multicolumn{1}{c|}{72\%} &
  28\% &
  \multicolumn{1}{c|}{72\%} &
  28\% &
  \multicolumn{1}{l|}{} &
  \multicolumn{1}{c|}{\textbf{5}} &
  \multicolumn{1}{l|}{Lord of the Flies} \\ \cline{1-11} \cline{13-14} 
\multicolumn{1}{|l|}{\textbf{Run 6}} &
  \multicolumn{1}{c|}{60\%} &
  40\% &
  \multicolumn{1}{c|}{56\%} &
  44\% &
  \multicolumn{1}{c|}{84\%} &
  16\% &
  \multicolumn{1}{c|}{86\%} &
  14\% &
  \multicolumn{1}{c|}{72\%} &
  28\% &
  &
  &
  \\ \cline{1-11}
\multicolumn{1}{|l|}{\textbf{Run 7}} &
  \multicolumn{1}{c|}{54\%} &
  46\% &
  \multicolumn{1}{c|}{50\%} &
  50\% &
  \multicolumn{1}{c|}{78\%} &
  22\% &
  \multicolumn{1}{c|}{64\%} &
  36\% &
  \multicolumn{1}{c|}{62\%} &
  38\% &
  &
  &
  \\ \cline{1-11}
\multicolumn{1}{|l|}{\textbf{Run 8}} &
  \multicolumn{1}{c|}{52\%} &
  48\% &
  \multicolumn{1}{c|}{40\%} &
  60\% &
  \multicolumn{1}{c|}{72\%} &
  28\% &
  \multicolumn{1}{c|}{68\%} &
  32\% &
  \multicolumn{1}{c|}{76\%} &
  24\% &
  &
  &
  \\ \cline{1-11}
\multicolumn{1}{|l|}{\textbf{Run 9}} &
  \multicolumn{1}{c|}{48\%} &
  52\% &
  \multicolumn{1}{c|}{52\%} &
  48\% &
  \multicolumn{1}{c|}{72\%} &
  28\% &
  \multicolumn{1}{c|}{74\%} &
  26\% &
  \multicolumn{1}{c|}{76\%} &
  24\% &
  &
  &
  \\ \cline{1-11}
\multicolumn{1}{|l|}{\textbf{Run 10}} &
  \multicolumn{1}{c|}{54\%} &
  46\% &
  \multicolumn{1}{c|}{52\%} &
  48\% &
  \multicolumn{1}{c|}{78\%} &
  22\% &
  \multicolumn{1}{c|}{76\%} &
  24\% &
  \multicolumn{1}{c|}{78\%} &
  22\% &
  &
  &
  \\ \cline{1-11}
\multicolumn{1}{|l|}{\textbf{Average}} &
  \multicolumn{1}{c|}{\textbf{52\%}} &
  48\% &
  \multicolumn{1}{c|}{\textbf{48\%}} &
  52\% &
  \multicolumn{1}{c|}{\textbf{76\%}} &
  24\% &
  \multicolumn{1}{c|}{\textbf{72\%}} &
  28\% &
  \multicolumn{1}{c|}{\textbf{75\%}} &
  25\% &
  &
  &
  \\ \cline{1-11}
\end{tabular}%
}
\end{table}


\begin{table}[]
\caption{The sentiment-level ranking of recommended movies for 2001: A Space Odyssey}
\label{tab:2001_space_sentiment}
\resizebox{\textwidth}{!}{%
\begin{tabular}{l|cc|cc|cc|cc|cc|lcl}
\cline{2-11}
\textbf{} &
  \multicolumn{2}{c|}{\textbf{Dark Star}} &
  \multicolumn{2}{c|}{\textbf{\begin{tabular}[c]{@{}c@{}}2010:   The Year We \\ Make Contact\end{tabular}}} &
  \multicolumn{2}{c|}{\textbf{Gravity}} &
  \multicolumn{2}{c|}{\textbf{The   Black Hole}} &
  \multicolumn{2}{c|}{\textbf{Ad   Astra}} &
  &
  \multicolumn{1}{l}{} &
  \\ \cline{2-11} \cline{13-14} 
\textbf{} &
  \multicolumn{1}{c|}{\textbf{Positive}} &
  \textbf{Negative} &
  \multicolumn{1}{c|}{\textbf{Positive}} &
  \textbf{Negative} &
  \multicolumn{1}{c|}{\textbf{Positive}} &
  \textbf{Negative} &
  \multicolumn{1}{c|}{\textbf{Positive}} &
  \textbf{Negative} &
  \multicolumn{1}{c|}{\textbf{Positive}} &
  \textbf{Negative} &
  \multicolumn{1}{l|}{} &
  \multicolumn{2}{c|}{\textbf{Order of Ranks}} \\ \cline{1-11} \cline{13-14} 
\multicolumn{1}{|l|}{Run 1} &
  \multicolumn{1}{c|}{64\%} &
  36\% &
  \multicolumn{1}{c|}{60\%} &
  40\% &
  \multicolumn{1}{c|}{58\%} &
  42\% &
  \multicolumn{1}{c|}{42\%} &
  58\% &
  \multicolumn{1}{c|}{34\%} &
  66\% &
  \multicolumn{1}{l|}{} &
  \multicolumn{1}{c|}{1} &
  \multicolumn{1}{l|}{2010:   The Year We Make Contact} \\ \cline{1-11} \cline{13-14} 
\multicolumn{1}{|l|}{Run 2} &
  \multicolumn{1}{c|}{70\%} &
  30\% &
  \multicolumn{1}{c|}{60\%} &
  40\% &
  \multicolumn{1}{c|}{66\%} &
  34\% &
  \multicolumn{1}{c|}{48\%} &
  52\% &
  \multicolumn{1}{c|}{40\%} &
  60\% &
  \multicolumn{1}{l|}{} &
  \multicolumn{1}{c|}{2} &
  \multicolumn{1}{l|}{Dark Star} \\ \cline{1-11} \cline{13-14} 
\multicolumn{1}{|l|}{Run 3} &
  \multicolumn{1}{c|}{56\%} &
  44\% &
  \multicolumn{1}{c|}{74\%} &
  26\% &
  \multicolumn{1}{c|}{54\%} &
  46\% &
  \multicolumn{1}{c|}{52\%} &
  48\% &
  \multicolumn{1}{c|}{22\%} &
  78\% &
  \multicolumn{1}{l|}{} &
  \multicolumn{1}{c|}{3} &
  \multicolumn{1}{l|}{Gravity} \\ \cline{1-11} \cline{13-14} 
\multicolumn{1}{|l|}{Run 4} &
  \multicolumn{1}{c|}{60\%} &
  40\% &
  \multicolumn{1}{c|}{62\%} &
  38\% &
  \multicolumn{1}{c|}{64\%} &
  36\% &
  \multicolumn{1}{c|}{46\%} &
  54\% &
  \multicolumn{1}{c|}{24\%} &
  76\% &
  \multicolumn{1}{l|}{} &
  \multicolumn{1}{c|}{4} &
  \multicolumn{1}{l|}{The Black Hole} \\ \cline{1-11} \cline{13-14} 
\multicolumn{1}{|l|}{Run 5} &
  \multicolumn{1}{c|}{74\%} &
  26\% &
  \multicolumn{1}{c|}{68\%} &
  32\% &
  \multicolumn{1}{c|}{52\%} &
  48\% &
  \multicolumn{1}{c|}{48\%} &
  52\% &
  \multicolumn{1}{c|}{30\%} &
  70\% &
  \multicolumn{1}{l|}{} &
  \multicolumn{1}{c|}{5} &
  \multicolumn{1}{l|}{Ad Astra} \\ \cline{1-11} \cline{13-14} 
\multicolumn{1}{|l|}{Run 6} &
  \multicolumn{1}{c|}{64\%} &
  36\% &
  \multicolumn{1}{c|}{70\%} &
  30\% &
  \multicolumn{1}{c|}{46\%} &
  54\% &
  \multicolumn{1}{c|}{52\%} &
  48\% &
  \multicolumn{1}{c|}{30\%} &
  70\% &
  &
  &
  \multicolumn{1}{c}{} \\ \cline{1-11}
\multicolumn{1}{|l|}{Run 7} &
  \multicolumn{1}{c|}{56\%} &
  44\% &
  \multicolumn{1}{c|}{56\%} &
  44\% &
  \multicolumn{1}{c|}{60\%} &
  40\% &
  \multicolumn{1}{c|}{46\%} &
  54\% &
  \multicolumn{1}{c|}{24\%} &
  76\% &
  &
  &
  \multicolumn{1}{c}{} \\ \cline{1-11}
\multicolumn{1}{|l|}{Run 8} &
  \multicolumn{1}{c|}{68\%} &
  32\% &
  \multicolumn{1}{c|}{64\%} &
  36\% &
  \multicolumn{1}{c|}{68\%} &
  32\% &
  \multicolumn{1}{c|}{48\%} &
  52\% &
  \multicolumn{1}{c|}{30\%} &
  70\% &
  &
  &
  \multicolumn{1}{c}{} \\ \cline{1-11}
\multicolumn{1}{|l|}{Run 9} &
  \multicolumn{1}{c|}{60\%} &
  40\% &
  \multicolumn{1}{c|}{68\%} &
  32\% &
  \multicolumn{1}{c|}{60\%} &
  40\% &
  \multicolumn{1}{c|}{56\%} &
  44\% &
  \multicolumn{1}{c|}{30\%} &
  70\% &
  &
  &
  \multicolumn{1}{c}{} \\ \cline{1-11}
\multicolumn{1}{|l|}{Run 10} &
  \multicolumn{1}{c|}{56\%} &
  44\% &
  \multicolumn{1}{c|}{64\%} &
  36\% &
  \multicolumn{1}{c|}{64\%} &
  36\% &
  \multicolumn{1}{c|}{44\%} &
  56\% &
  \multicolumn{1}{c|}{32\%} &
  68\% &
  &
  &
  \multicolumn{1}{c}{} \\ \cline{1-11}
\multicolumn{1}{|l|}{\textbf{Average}} &
  \multicolumn{1}{c|}{\textbf{63\%}} &
  37\% &
  \multicolumn{1}{c|}{\textbf{65\%}} &
  35\% &
  \multicolumn{1}{c|}{\textbf{59\%}} &
  41\% &
  \multicolumn{1}{c|}{\textbf{48\%}} &
  52\% &
  \multicolumn{1}{c|}{\textbf{30\%}} &
  70\% &
  &
  &
  \multicolumn{1}{c}{} \\ \cline{1-11}
\end{tabular}%
}
\end{table}

Compiling all the results, we note the following Sentiment scores for the recommended movies in Table \ref{tab:final_sentiment_score}.
\begin{table}[]
\caption{The positivity score of all recommended movies based on the sentiment analysis algorithm}
\label{tab:final_sentiment_score}
\resizebox{\textwidth}{!}{%
\begin{tabular}{|lclclc|}
\hline
\multicolumn{6}{|c|}{\textbf{Sentiment Analysis of Recommended Movies}}                                                           \\ \hline
\multicolumn{2}{|c|}{\textbf{Tenet}} & \multicolumn{2}{c|}{\textbf{CastAway}} & \multicolumn{2}{c|}{\textbf{2001: Space Odyssey}} \\ \hline
\multicolumn{1}{|l|}{Predestination} &
  \multicolumn{1}{c|}{\textbf{0.726}} &
  \multicolumn{1}{l|}{Nim's Island} &
  \multicolumn{1}{c|}{\textbf{0.758}} &
  \multicolumn{1}{l|}{2010: The Year We Make Contact} &
  \textbf{0.646} \\ \hline
\multicolumn{1}{|l|}{Interstellar} &
  \multicolumn{1}{c|}{\textbf{0.668}} &
  \multicolumn{1}{l|}{The Most Dangerous Game} &
  \multicolumn{1}{c|}{\textbf{0.746}} &
  \multicolumn{1}{l|}{Dark Star} &
  \textbf{0.628} \\ \hline
\multicolumn{1}{|l|}{Mission Impossible: Fallout} &
  \multicolumn{1}{c|}{\textbf{0.644}} &
  \multicolumn{1}{l|}{The Blue Lagoon} &
  \multicolumn{1}{c|}{\textbf{0.724}} &
  \multicolumn{1}{l|}{Gravity} &
  \textbf{0.592} \\ \hline
\multicolumn{1}{|l|}{The Man From U.N.C.L.E} &
  \multicolumn{1}{c|}{\textbf{0.64}} &
  \multicolumn{1}{l|}{Six Days Seven Nights} &
  \multicolumn{1}{c|}{\textbf{0.52}} &
  \multicolumn{1}{l|}{The Black Hole} &
  \textbf{0.482} \\ \hline
\multicolumn{1}{|l|}{The 355} &
  \multicolumn{1}{c|}{\textbf{0.452}} &
  \multicolumn{1}{l|}{Lord of the Flies} &
  \multicolumn{1}{c|}{\textbf{0.478}} &
  \multicolumn{1}{l|}{Ad Astra} &
  \textbf{0.296} \\ \hline
\end{tabular}%
}
\end{table}

\subsection{Movie Similarity Score}
The final movie similarity scores are listed in Table \ref{tab:movie_similarity_score}. These score are derived from the weighted sum methodology explained in Section \ref{ssec:movie_similarity_calculation} 

\begin{table}[]
\caption{The Movie Similarity Score based on the weighted combination of Sentiment Score and Video Similarity Score}
\label{tab:movie_similarity_score}
\resizebox{\textwidth}{!}{%
\begin{tabular}{c|lc|l}
\cline{2-3}
\multicolumn{1}{l|}{}                                    & \multicolumn{2}{c|}{\textbf{Proposed Ranking Algorithm}}  &  \\ \cline{2-3}
\multicolumn{1}{l|}{} &
  \multicolumn{1}{l|}{\textbf{Recommended Movie}} &
  \multicolumn{1}{l|}{\textbf{Movie Similarity Score}} &
   \\ \cline{1-3}
\multicolumn{1}{|c|}{\multirow{5}{*}{\textbf{Tenet}}}    & \multicolumn{1}{l|}{Predestination}              & 0.809  &  \\ \cline{2-3}
\multicolumn{1}{|c|}{}                                   & \multicolumn{1}{l|}{Interstellar}                & 0.7405 &  \\ \cline{2-3}
\multicolumn{1}{|c|}{}                                   & \multicolumn{1}{l|}{Mission Impossible: Fallout} & 0.7885 &  \\ \cline{2-3}
\multicolumn{1}{|c|}{}                                   & \multicolumn{1}{l|}{The Man From U.N.C.L.E}      & 0.7595 &  \\ \cline{2-3}
\multicolumn{1}{|c|}{}                                   & \multicolumn{1}{l|}{The 355}                     & 0.6385 &  \\ \cline{1-3}
\multicolumn{1}{|c|}{\multirow{5}{*}{\textbf{CastAway}}} & \multicolumn{1}{l|}{Nim's Island}                & 0.8145 &  \\ \cline{2-3}
\multicolumn{1}{|c|}{}                                   & \multicolumn{1}{l|}{The Most Dangerous Game}     & 0.7755 &  \\ \cline{2-3}
\multicolumn{1}{|c|}{}                                   & \multicolumn{1}{l|}{The Blue Lagoon}             & 0.739  &  \\ \cline{2-3}
\multicolumn{1}{|c|}{}                                   & \multicolumn{1}{l|}{Six Days Seven Nights}       & 0.651  &  \\ \cline{2-3}
\multicolumn{1}{|c|}{}                                   & \multicolumn{1}{l|}{Lord of the Flies}           & 0.621  &  \\ \cline{1-3}
\multicolumn{1}{|c|}{\multirow{5}{*}{\textbf{2001: Space Odyssey}}} &
  \multicolumn{1}{l|}{2010: The Year We Make Contact} &
  0.666 &
   \\ \cline{2-3}
\multicolumn{1}{|c|}{}                                   & \multicolumn{1}{l|}{Dark Star}                   & 0.6835 &  \\ \cline{2-3}
\multicolumn{1}{|c|}{}                                   & \multicolumn{1}{l|}{Gravity}                     & 0.6975 &  \\ \cline{2-3}
\multicolumn{1}{|c|}{}                                   & \multicolumn{1}{l|}{The Black Hole}              & 0.6075 &  \\ \cline{2-3}
\multicolumn{1}{|c|}{}                                   & \multicolumn{1}{l|}{Ad Astra}                    & 0.5335 &  \\ \cline{1-3}
\end{tabular}%
}
\end{table}

\subsection{Comparative Study with Existing System}
To compare our rankings, we shall use the publicly available IMDb Top 250 ranking algorithm along with another publicly available metric of the popularity of the movie to rank the recommendations and then compare the results. However, we extend this algorithm to all the movies as our dataset has over 10,000 movies. The IMDb ranking algorithm is as follows 

\begin{equation}
    w = \frac{(r \times v) +(c \times m)}{v + m}
\end{equation}
where,

\begin{center}
    w is the weighted rating \\
    r is the average rating of the movie \\
    v is the number of ratings for the movie \\
    c is the mean of the ratings of all the movies in the corpus \\
    m is the minimum votes required to be listed
\end{center}

We choose to take the inverse of popularity because lower the value of popularity, more popular the movie is. The final score is based on

\begin{equation}
    Final \ Score = (W \times 0.5) + (\frac{1}{P} \times 0.5)
\end{equation}

where, W is normalized weighted rating and P is normalized popularity \\

Note: The results of the comparative study are listed in Table \ref{tab:comparison}. However, due to the subjective nature of recommendations and lack of theoretical validation on what type of recommendation is better than the other, we do not utilize any performance metric to determine what type of system is better than the other.

\begin{table}[]
\caption{The comparison of our proposed methodology with the ranking of IMDBs algorithm}
\label{tab:comparison}
\resizebox{\textwidth}{!}{%
\begin{tabular}{c|c|c|c|}
\cline{2-4}
\multicolumn{1}{l|}{} &
  \multicolumn{1}{l|}{\textbf{Rank}} &
  \multicolumn{1}{l|}{\textbf{Proposed Ranking Algorithm}} &
  \multicolumn{1}{l|}{\textbf{\begin{tabular}[c]{@{}l@{}}Current Ranking Algorithm based\\  on Ratings and Popularity\end{tabular}}} \\ \hline
\multicolumn{1}{|c|}{\multirow{5}{*}{\textbf{Tenet}}}               & \textbf{1} & Predestination                 & Interstellar                  \\ \cline{2-4} 
\multicolumn{1}{|c|}{}                                              & \textbf{2} & Mission Impossible: Fallout    & Mission Impossible: Fallout   \\ \cline{2-4} 
\multicolumn{1}{|c|}{}                                              & \textbf{3} & The Man From U.N.C.L.E         & Predestination                \\ \cline{2-4} 
\multicolumn{1}{|c|}{}                                              & \textbf{4} & Interstellar                   & The Man from U.N.C.L.E        \\ \cline{2-4} 
\multicolumn{1}{|c|}{}                                              & \textbf{5} & The 355                        & The 355                       \\ \hline
\multicolumn{1}{|c|}{\multirow{5}{*}{\textbf{CastAway}}}            & \textbf{1} & Nim's Island                   & The Most Dangerous Game       \\ \cline{2-4} 
\multicolumn{1}{|c|}{}                                              & \textbf{2} & The Most Dangerous Game        & Lord of The Flies             \\ \cline{2-4} 
\multicolumn{1}{|c|}{}                                              & \textbf{3} & The Blue Lagoon                & Nim's Island                  \\ \cline{2-4} 
\multicolumn{1}{|c|}{}                                              & \textbf{4} & Six Days Seven Nights          & Blue Lagoon                   \\ \cline{2-4} 
\multicolumn{1}{|c|}{}                                              & \textbf{5} & Lord of the Flies              & Six Days Seven Nights         \\ \hline
\multicolumn{1}{|c|}{\multirow{5}{*}{\textbf{2001: Space Odyssey}}} & \textbf{1} & Gravity                        & Gravity                       \\ \cline{2-4} 
\multicolumn{1}{|c|}{}                                              & \textbf{2} & Dark Star                      & 2010: The Year We Make Contact\\ \cline{2-4} 
\multicolumn{1}{|c|}{}                                              & \textbf{3} & 2010: The Year We Make Contact & Ad Astra                      \\ \cline{2-4} 
\multicolumn{1}{|c|}{}                                              & \textbf{4} & The Black Hole                 & Dark Star                     \\ \cline{2-4} 
\multicolumn{1}{|c|}{}                                              & \textbf{5} & Ad Astra                       & The Black Hole                \\ \hline
\end{tabular}%
}
\end{table}

\section{Conclusion and Future Work}
To conclude, our contribution proposes a content-based recommendation system
for movies enhanced with a ranking algorithm that considers the visual similarity of
the content itself as well as measures the sentiment of user reviews. For visual
similarity, we use a pre-trained VGG network for feature extraction from key
frames of the movie trailer. We follow this by clustering and calculating similarity based on the Euclidean distance of the distribution of frames of the test and the reference movie. For sentiment analysis, we utilize a publicly available IMDb dataset and choose the model with the best combination of Accuracy and F1 score followed by calculating the \% of positive reviews in the movie. Finally, we combine these scores to create a unified Movie Similarity Score. We
then compare our results with the ranking algorithm of IMDb and see that the
results are noticeably different. 

We believe our contribution can improve the quality of recommendations compared to the existing content-based systems. Our methodology is not only limited to ranking content-based recommendations based on visual similarity and audience sentiment, but also
other dimensions such as the auditory similarity and closeness of the script. Our
contribution can also be used to recommend other types of visual media, such
as YouTube videos and animated content. It can also be used as an add-on to
existing recommendation systems to incorporate a "similar feel" factor in the
recommendations. To take this research further, we plan to design a survey that tries to curate the opinion of movie watchers regarding the quality of our recommendations. This would act as a validation metric, allowing us to measure the improvement in recommendations compared to a traditional content-based system.

%
%

\bibliographystyle{splncs04}
\bibliography{mybibliography}

\begin{thebibliography}{8}
\bibitem{1-intro}
Sharma R, Singh R. Evolution of recommender systems from ancient times to modern era: a survey. Indian Journal of Science and Technology. 2016 May;9(20):1-2.

\bibitem{2-intro}
Isinkaye FO, Folajimi YO, Ojokoh BA. Recommendation systems: Principles, methods and evaluation. Egyptian informatics journal. 2015 Nov 1;16(3):261-73.


\bibitem{2-lit}
Goldberg D, Nichols D, Oki BM, Terry D. Using collaborative filtering to weave an information tapestry. Communications of the ACM. 1992 Dec 1;35(12):61-70.



\bibitem{24-lit}
Rashid AM, Albert I, Cosley D, Lam SK, McNee SM, Konstan JA, Riedl J. Getting to know you: learning new user preferences in recommender systems. InProceedings of the 7th international conference on Intelligent user interfaces 2002 Jan 13 (pp. 127-134).

\bibitem{25-lit}
Resnick P, Varian HR. Recommender systems. Communications of the ACM. 1997 Mar 1;40(3):56-8.

\bibitem{9-lit}
Song B, Gao Y, Li XM. Research on collaborative filtering recommendation algorithm based on mahout and user model. InJournal of Physics: Conference Series 2020 (Vol. 1437, No. 1, p. 012095). IOP Publishing.

\bibitem{10-lit}
Linden G, Smith B, York J. Amazon. com recommendations: Item-to-item collaborative filtering. IEEE Internet computing. 2003 Jan 22;7(1):76-80.

\bibitem{11-lit}
Philip S, Shola P, Ovye A. Application of content-based approach in research paper recommendation system for a digital library. International Journal of Advanced Computer Science and Applications. 2014 Oct;5(10).

\bibitem{26-lit}
Van Meteren R, Van Someren M. Using content-based filtering for recommendation. InProceedings of the machine learning in the new information age: MLnet/ECML2000 workshop 2000 May 30 (Vol. 30, pp. 47-56).

\bibitem{12-lit}
Adomavicius G, Tuzhilin A. Toward the next generation of recommender systems: A survey of the state-of-the-art and possible extensions. IEEE transactions on knowledge and data engineering. 2005 Apr 25;17(6):734-49.

\bibitem{13-lit}
Göksedef M, Gündüz-Öğüdücü Ş. Combination of Web page recommender systems. Expert Systems with Applications. 2010 Apr 1;37(4):2911-22.


\bibitem{15-lit}
Baidada M, Mansouri K, Poirier F. Hybrid Filtering Recommendation System in an Educational Context: Experiment in Higher Education in Morocco. International Journal of Web-Based Learning and Teaching Technologies (IJWLTT). 2022 Jan 1;17(1):1-7.

\bibitem{27-lit}
Çano E, Morisio M. Hybrid recommender systems: A systematic literature review. Intelligent Data Analysis. 2017 Jan 1;21(6):1487-524.

\bibitem{28-lit}
Dong Y, Liu S, Chai J. Research of hybrid collaborative filtering algorithm based on news recommendation. In2016 9th international congress on image and signal processing, biomedical engineering and informatics (CISP-BMEI) 2016 Oct 15 (pp. 898-902). IEEE.

\bibitem{29-lit}
Konstas I, Stathopoulos V, Jose JM. On social networks and collaborative recommendation. InProceedings of the 32nd international ACM SIGIR conference on Research and development in information retrieval 2009 Jul 19 (pp. 195-202).

\bibitem{20-lit}
Deldjoo Y, Elahi M, Cremonesi P, Garzotto F, Piazzolla P, Quadrana M. Content-based video recommendation system based on stylistic visual features. Journal on Data Semantics. 2016 Jun;5(2):99-113.

\bibitem{21-lit}
Kvifte T, Elahi M, Trattner C. Hybrid Recommendation of Movies Based on Deep Content Features. InInternational Conference on Service-Oriented Computing 2022 (pp. 32-45). Springer, Cham.


\bibitem{22-lit}
Asani E, Vahdat-Nejad H, Sadri J. Restaurant recommender system based on sentiment analysis. Machine Learning with Applications. 2021 Dec 15;6:100114.

\bibitem{23-lit}
Chauhan A, Nagar D, Chaudhary P. Movie Recommender system using Sentiment Analysis. In2021 International Conference on Innovative Practices in Technology and Management (ICIPTM) 2021 Feb 17 (pp. 190-193). IEEE.




\bibitem{imdb_ref}
IMDb Homepage, \url{https://www.imdb.com/}. Last accessed 27 Sept 2022

\bibitem{youtube_ref}
YouTube Website, \url{https://www.youtube.com/}. Last accessed 27 Nov 2022




\bibitem{1-methodContent}
Singla R, Gupta S, Gupta A, Vishwakarma DK. FLEX: a content based movie recommender. In2020 International Conference for Emerging Technology (INCET) 2020 Jun 5 (pp. 1-4). IEEE.

\bibitem{2-methodContent}
Jones KS. A statistical interpretation of term specificity and its application in retrieval. Journal of documentation. 1972.

\bibitem{3-methodContent}
Robertson S. Understanding inverse document frequency: on theoretical arguments for IDF. Journal of documentation. 2004 Oct 1.

\bibitem{4-methodContent}
Gunawan D, Sembiring CA, Budiman MA. The implementation of cosine similarity to calculate text relevance between two documents. InJournal of physics: conference series 2018 Mar 1 (Vol. 978, No. 1, p. 012120). IOP Publishing.

\bibitem{1-methodSentiment}
Mishne G, Glance NS. Predicting movie sales from blogger sentiment. InAAAI spring symposium: computational approaches to analyzing weblogs 2006 Mar 27 (pp. 155-158).

\bibitem{2-methodSentiment}
Cox DR. The regression analysis of binary sequences. Journal of the Royal Statistical Society: Series B (Methodological). 1958 Jul;20(2):215-32.

\bibitem{3-methodSentiment}
Quinlan JR. Induction of decision trees. Machine learning. 1986 Mar;1(1):81-106.

\bibitem{4-methodSentiment}
Ho TK. Random decision forests. InProceedings of 3rd international conference on document analysis and recognition 1995 Aug 14 (Vol. 1, pp. 278-282). IEEE.



\bibitem{5-methodSentiment}
Chen T, Guestrin C. Xgboost: A scalable tree boosting system. InProceedings of the 22nd acm sigkdd international conference on knowledge discovery and data mining 2016 Aug 13 (pp. 785-794).

\bibitem{6-methodSentiment}
Webb GI, Keogh E, Miikkulainen R. Naïve Bayes. Encyclopedia of machine learning. 2010;15:713-4.

\bibitem{7-methodSentiment}
Yasen M, Tedmori S. Movies reviews sentiment analysis and classification. In2019 IEEE Jordan International Joint Conference on Electrical Engineering and Information Technology (JEEIT) 2019 Apr 9 (pp. 860-865). IEEE.


\bibitem{1-methodVideo}
Ouyang S, Zhong L, Luo R. The comparison and analysis of extracting video key frame. InIOP Conference Series: Materials Science and Engineering 2018 May 1 (Vol. 359, No. 1, p. 012010). IOP Publishing.



\bibitem{4-methodVideo}
Karray S, Debernitz L. The effectiveness of movie trailer advertising. International Journal of Advertising. 2017 Mar 4;36(2):368-92.

\bibitem{5-methodVideo}
Castellano, B.: Pyscenedetect. http://github.com/Breakthrough/PySceneDetect(07 2012)

\bibitem{6-methodVideo}
Simonyan K, Zisserman A. Very deep convolutional networks for large-scale image recognition. arXiv preprint arXiv:1409.1556. 2014 Sep 4.


\bibitem{8-methodVideo}
Ahmed T, Das P, Ali MF, Mahmud MF. A comparative study on convolutional neural network based face recognition. In2020 11th International Conference on Computing, Communication and Networking Technologies (ICCCNT) 2020 Jul 1 (pp. 1-5). IEEE.

\bibitem{9-methodVideo}
Bansal R, Raj G, Choudhury T. Blur image detection using Laplacian operator and Open-CV. In2016 International Conference System Modeling \& Advancement in Research Trends (SMART) 2016 Nov 25 (pp. 63-67). IEEE.

\bibitem{10-methodVideo}
Na S, Xumin L, Yong G. Research on k-means clustering algorithm: An improved k-means clustering algorithm. In2010 Third International Symposium on intelligent information technology and security informatics 2010 Apr 2 (pp. 63-67). Ieee.

\bibitem{11-methodVideo}
Shi C, Wei B, Wei S, Wang W, Liu H, Liu J. A quantitative discriminant method of elbow point for the optimal number of clusters in clustering algorithm. EURASIP Journal on Wireless Communications and Networking. 2021 Dec;2021(1):1-6.

\bibitem{imdbDS}
Maas A, Daly RE, Pham PT, Huang D, Ng AY, Potts C. Learning word vectors for sentiment analysis. InProceedings of the 49th annual meeting of the association for computational linguistics: Human language technologies 2011 Jun (pp. 142-150).

\bibitem{tenet}
Tenet. [Film] Directed by: Christopher Nolan. United States: Warner Bros.; 2020.

\bibitem{castaway}
Cast Away. [Film] Directed by: Robert Zemeckis. United States: Twentieth Century Fox; 2000.

\bibitem{2001spaceodyssey}
2001: A Space Odyssey. [Film] Directed by: Stanley Kubrick. United Kingdom: Metro-Goldwyn-Mayer (MGM), Stanley Kubrick Productions; 1968.

\bibitem{interstellar}
Interstellar. [Film] Directed by: Christopher Nolan. United States: Paramount Pictures; 2014.

\bibitem{the355}
The 355. [Film] Directed by: Simon Kinberg. United States: Universal Pictures; 2022.

\bibitem{predestination}
Predestination. [Film] Directed by: Michael Spierig, Peter Spierig. Australia: Screen Australia; 2014.

\bibitem{manfromuncle}
The Man from U.N.C.L.E. [Film] Directed by: Guy Ritchie. United States: Warner Bros; 2015.


\bibitem{mifallout}
Mission: Impossible - Fallout. [Film] Directed by: Christopher McQuarrie. United States: Paramount Pictures; 2018.



\bibitem{6days7nights}
Six Days Seven Nights. [Film] Directed by: Ivan Reitman. United States: Touchstone Pictures; 1998.

\bibitem{lordoftheflies}
Lord of the Flies. [Film] Directed by: Harry Hook. United States: Castle Rock Entertainment; 1990.

\bibitem{nimsisland}
Nim's Island. [Film] Directed by: Jennifer Flackett, Mark Levin. United States: Walden Media; 2008.


\bibitem{bluelagoon}
The Blue Lagoon. [Film] Directed by: Randal Kleiser. United States: Columbia Pictures; 1980.

\bibitem{mostdangerousgame}
The Most Dangerous Game. [Film] Directed by: Irving Pichel, Ernest B. Schoedsack. United States: Merian C. Cooper and Ernest Schoedsack; 1932.

\bibitem{darkstar}
Dark Star. [Film] Directed by: John Carpenter. United States: Jack H. Harris Enterprises, University of Southern California (USC); 1974.

\bibitem{2010yearwemakecontact}
2010: The Year We Make Contact. [Film] Directed by: Peter Hyams. United States: Metro-Goldwyn-Mayer (MGM); 1984.

\bibitem{gravity}
Gravity. [Film] Directed by: Alfonso Cuarón. United Kingdom: Warner Bros. ; 2013.


\bibitem{blackhole}
The Black Hole. [Film] Directed by: Gary Nelson. United States: Walt Disney Productions; 1974.

\bibitem{adastra}
Ad Astra. [Film] Directed by: James Gray. United States: New Regency Productions; 1974.




\end{thebibliography}

\end{document}